\documentclass[conference]{IEEEtran}
\IEEEoverridecommandlockouts
\usepackage{cite}
\usepackage{amsmath,amssymb,amsfonts}
\usepackage{bm}

\usepackage{graphicx}
\usepackage{textcomp}
\usepackage{xcolor}
\usepackage{verbatim} 
\usepackage{soul}
\usepackage{booktabs}
\usepackage{hyperref}
\usepackage{algorithm}
\usepackage[noend]{algpseudocode}

\usepackage{array,multirow} 
\usepackage{makecell}

\newcommand{\mcc}[1]{\multicolumn{1}{c}{#1}}

\def\BibTeX{{\rm B\kern-.05em{\sc i\kern-.025em b}\kern-.08em
    T\kern-.1667em\lower.7ex\hbox{E}\kern-.125emX}}
\begin{document}

\title{Using Graph Theory for Improving Machine Learning-based Detection of Cyber Attacks

\thanks{This work was supported in part by the GiBIDS (Graph-Based Intrusion Detection System) research project, funded by GÉANT under the GÉANT Innovation Programme.}
}

\author{\IEEEauthorblockN{Giacomo Zonneveld, Lorenzo Principi, Marco Baldi}
\IEEEauthorblockA{\textit{Department of Information Engineering}\\
\textit{Università Politecnica delle Marche}\\
Ancona, Italy \\
}}

\maketitle

\begin{abstract}
Early detection of network intrusions and cyber threats is one of the main pillars of cybersecurity. One of the most effective approaches for this purpose is to analyze network traffic with the help of artificial intelligence algorithms, with the aim of detecting the possible presence of an attacker by distinguishing it from a legitimate user.
This is commonly done by collecting the traffic exchanged between terminals in a network and analyzing it on a per-packet or per-connection basis.
In this paper, we propose instead to perform pre-processing of network traffic under analysis with the aim of extracting some new metrics on which we can perform more efficient detection and overcome some limitations of classical approaches.
These new metrics are based on graph theory, and consider the network as a whole, rather than focusing on individual packets or connections.
Our approach is validated through experiments performed on publicly available data sets, from which it results that it can not only overcome some of the limitations of classical approaches, but also achieve a better detection capability of cyber threats.
\end{abstract}


\section{Introduction}
We study network intrusion detection techniques based on the analysis of network traffic through machine learning techniques, which represent one of the most promising techniques today for promptly detecting cyber threats and thus increasing cybersecurity.
According to classical approaches, intrusion detection based on the analysis of network traffic focuses on network protocol characteristics like:
\begin{itemize}
    \item network layer protocol parameters (IP addresses, etc.) \cite{Abdulhammed, Thockchom},
    \item transport layer protocol parameters (TCP/UDP port, TCP session \cite{rodriguez}, etc.) \cite{Abdulhammed,Thockchom},
    \item traffic-related features (number of packet transmissions, etc.) \cite{Abdulhammed,rodriguez,Thockchom},
    \item content-related features (payload length, TCP flags, etc.) \cite{Abdulhammed,rodriguez,Thockchom},
    \item time-related features (connection duration, etc.) \cite{Abdulhammed,rodriguez,Thockchom},
    \item DNS-based features, since malicious P2P networks use DNS to establish connections \cite{6362965}.
\end{itemize}

Machine learning (ML) techniques are then used to learn different behaviours by analyzing these network features in order to be able to detect the presence of anomalies.

The main limitations in using these features are that: i) they have loose correlation with the actual presence of a malware, ii) each terminal is considered individually, rather than looking at the network as a whole, iii) attackers could camouflage their behavior by using evasive techniques such as packet manipulation to hinder detection and iv) the use of secure protocols exploiting encrypted payloads limits the ability to analyse packets.

In this work, we propose an alternative approach relying on aggregated network features derived from graph theory rather than using the aforementioned specific network features.
More in detail, we propose to perform detection based on a graphical representation of the network, which only uses information concerning the identity of connected terminals (e.g. their IP address) and the times each connection occurs.

We validate our approach through numerical simulations considering the openly available CIC-IDS2017 data set \cite{CIC}, and we show that the graph-based detection approach we propose is able to outperform classical approaches relying on network protocol features.

Several works exist in the literature on the use of ML techniques for network intrusion detection, in most cases exploiting classical network protocol features.

In \cite{Abdulhammed} classification results obtained by selecting features through an auto-encoder and PCA are comparatively assessed.
In \cite{rodriguez} a combination of correlation analysis with an exhaustive search is exploited to find the best subset of features. 
Both \cite{Abdulhammed} and \cite{rodriguez} use a Random Forest classifier, with the approach in \cite{Abdulhammed} achieving better results than that in \cite{rodriguez}.
In \cite{Thockchom} feature selection based on the chi-square test is proposed and an ensemble classifier is adopted, showing that applying such an ensemble classifier to selected features through the chi-square test provides good classification performance.

Compared with these previous works, which use traditional protocol features, our approach is novel in that it uses other features, derived through the use of graph theory concepts.
Graphs are not new in this context and in fact they have already been used to represent interactions between network endpoints. For example, in \cite{dss} graph-based metrics (also considered in this work) are used to decide where security tools such as firewalls and IDS should be installed in a complex network. 
In \cite{vuln-detection}, such metrics are used to identify most vulnerable systems within a network, in particular by considering edges generation. In \cite{graph-decomp}, structural changes in graphs are used to determine whether an attack has occurred or not.
This is done on a daily basis, by applying clustering techniques and analyzing the overall shape of daily graphs.
In \cite{mal-netminter} and \cite{mal-network}, it is proposed to use ML algorithms to classify different types of malware by generating graph-based features extracted from a graph representing system calls made by such malware.

Differently from these previous works, we use graph theory concepts to extract a set of new features from network traffic data sets, and we use such new features for supervised ML-based detection of single malicious network connections.

The paper is organized as follows: 
in Section \ref{sec:background} we introduce some background notions and the notation we use, in Section \ref{sec:datasetgeneration} we describe how the graph-based features we consider are extracted from network traffic data sets, in Section \ref{sec:classification} we introduce and tune ML-based classifiers for analyzing the data set we consider, in Section \ref{sec:results} we validate our approach through numerical simulations and in Section \ref{sec:conclusion} we draw some conclusive remarks.

\section{Background, tools and data sets
\label{sec:background}
}

A graph \(G = (V, E)\) is a pair of a finite set of nodes \(V=\{v_1,...,v_n\}\) and a finite set of edges \(E=\{(v_i,v_j) \:\big|\: 1\leq i,j \leq n\}\), which represent a link between two nodes.
If the graph is oriented, then each edge can be traversed in only one-way, going from $v_i$ to $v_j$, and is defined by an \textit{ordered} pair of vertices $(v_i, v_j)$, meaning that $(v_j, v_i)$ will represent another edge. Otherwise, each edge is defined by an \textit{unordered} pair of vertices. Furthermore, to each edge we can associate a weight, i.e., a numerical value used and differently interpreted by the metric algorithms defined later.
We denote by $e=\{(v_i, v_j), w \}$ a generic edge linking $v_i, v_j \in V$ where $w$ is the associated weight, since that we use oriented graphs\footnote{Although some metrics require a non-oriented graph, since switching from an oriented to a non-oriented graph is straightforward (and not the reverse), each edge will be considered oriented.}.

\begin{figure}[ht]
    \centering
    \includegraphics[width=0.4\textwidth]{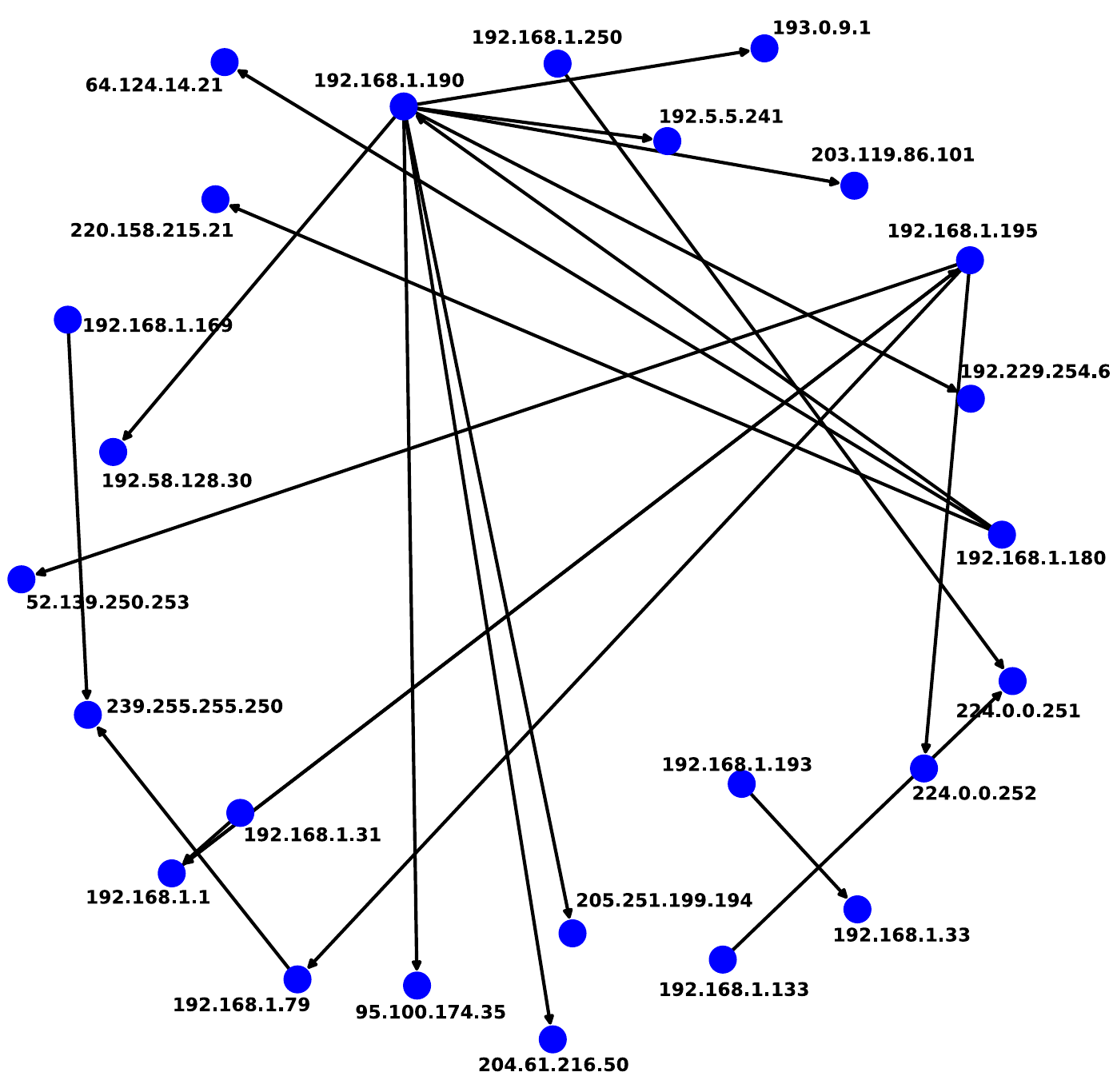}
    \caption{Example of graph representing network traffic}
    \label{fig:bicolor_graph}
\end{figure}

In our work, each network terminal provided with an IP address is considered as a node in a graph, while edges are the occurred connections between any two terminals, oriented from source to destination.
This way, we obtain graphs of the type reported in Fig. \ref{fig:bicolor_graph}.

\subsection{Graph-based metrics}

Given a populated graph $G=(V,E)$, for each node $v \in V$ we can calculate some classical graph theory metrics \cite{graph-metrics}, namely:
\begin{itemize}
    \item The \textit{Degree centrality} (\textit{DC}) and its variants expresses how much the node is directly connected with all the other nodes. If the graph is oriented, this metric can be evaluated for outgoing edges with \textit{Out-Degree centrality}, and for incoming edges with \textit{In-Degree centrality}.
    \item The \textit{Closeness centrality} considers how much the node is near all the other nodes.
    \item The \textit{Betweenness centrality} (\textit{BC}) considers how many times the node is a crossing point in all the shortest paths evaluated for each couple of nodes.
    \item The \textit{Eigenvector centrality} evaluates the extent to which the given node is directly connected to nodes that are themselves strongly connected to the rest of the network.
    \item The \textit{Clustering coefficient} (\textit{CC}) measures for the given node how much nearby nodes are directly connected to each other. For its evaluation we consider the formulation defined in \cite{c_coeff} and implemented in \cite{graph-tool}. Such formulation allows evaluating the measure taking nodes at a chosen distance from the given node instead of only directly connected nodes. The considered distances are \(d=1\) (default) and \(d=2\).
\end{itemize}
The aforementioned metrics are summarized in Table \ref{tab:features}, along with their variability domains and weight interpretation.

\begin{table}
    \centering
    \caption{Graph-based metrics}
    \begin{tabular}{lll}\toprule
        \textbf{Metric name} & \textbf{Domain} & \textbf{Weight effect} \\\midrule
        Degree centrality & \(\mathbb{N}^+\) & Connection strength\\
        In-Degree centrality & \(\mathbb{N}\) & Connection strength\\
        Out-Degree centrality & \(\mathbb{N}\) & Connection strength\\
        Closeness centrality & \([0,1]{\cup}\{\text{-}10\}\)\footnotemark & Traversal cost\\
        Betweenness centrality & \([0,1]{\cup}\{\text{-}10\}\)\footnotemark[\value{footnote}] & Traversal cost\\
        Eigenvector centrality & \([0,1]{\cup}\{\text{-}10\}\)\footnotemark[\value{footnote}] & Traversal cost\\
        Clustering coefficient d=1 & \([0,1]\) & No weights\\
        Clustering coefficient d=2 & \([0,1]\) & No weights \\\bottomrule
    \end{tabular}
    \label{tab:features}
\end{table}
\footnotetext{Infinite values have been transformed into \(-10\), which is an out-of-range value, but other choices are possible.}

\subsection{ML-based classification}

We use the metrics introduced above to feed ML classification algorithms based on the so-called Support Vector Machine (SVM), introduced in \cite{svm}. Such an algorithm, by mapping the input data into a higher dimensional feature space through the application of a chosen non-linear function, can solve classification problems by defining a linear decision surface capable of separating data instances belonging to different classes. This is done by selecting certain data instances, called \textit{support vectors}, that are close to the decision surface. The non-linear function we consider in this work is the well-known Radial Basis Function, defined as
\begin{equation}
    RBF(x,y) = e^{-\gamma||x-y||^2},
\end{equation}
where \(\gamma\) is a coefficient that controls the impact of each training instance on the definition of the decision surface. The higher its value, the less regular the surface will be.
A penalty parameter \(C\) is used to adjust for misclassification of training samples due to a non-perfectly definable separation surface. Large values of $C$ force the classifier to separate training instances correctly, but reduce generality, leading to possible overfitting problems when classifying test instances.

\section{Graph-based data set generation
\label{sec:datasetgeneration}
}

In this section we explain how graph-based features can be extracted from network traffic data sets.
In particular, we consider labeled data sets containing lists of connections.

By \textit{connection} we mean an information record representing the summary of a communication happened between two network terminals using some network protocol\footnote{Which network packets are included in a connection is obvious for those protocols where the beginning and end of the communication are explicit (such as TCP), otherwise time boundaries are used.}. %
Each connection is also labelled as \textit{benign} or \textit{malicious}, in the latter case also including the attack name, when available.
More in detail, the generic $i$-th connection $c_i$ is defined by:
\begin{itemize}
    \item $z^s_i$ and $z^d_i$, the source and the destination IP addresses, respectively.
    \item $t_i$, a timestamp corresponding to transmission of the first packet.
    \item $m_i$, a parameter indicating if the connection is benign or malicious.
    \item \(\bm{a}_i = \{a_{i_{0}},...,a_{i_{n^a \text{-} 1}}\}\), a set of $n^a$ metrics calculated by the network tool considering all the packets belonging to the $i$-th connection (as stated above, we refer to these as the \textit{classic features}).
\end{itemize}

We can hence define a data set
\[
D = [ c_0, ..., c_{N-1} ]
\]
as an array of chronologically ordered connections, with 
\[
c_i = \{ z^s_i, z^d_i, t_i, m_i, \bm{a}_i \}.
\]

\subsection{Graph population}

Given a graph $G=(V, E)$ in which we want to insert a new edge $e = (v_a, v_b)$, we define the updating function
\begin{equation*}
    \Lambda(G, e) = G^+ = (V^+,E^+),
\end{equation*}
where
\begin{align*}
    V^+ &= V \cup \{ v_a \}  \cup \{ v_b \} \\
    E^+ &= \varepsilon(E, (v_a, v_b), \omega)
\end{align*}
and $\varepsilon$ is a function that depends on the parameter $\omega$ and works as follows:
\begin{itemize}
    \item $\omega = \omega^u$ (unweighted case): add a new edge $e=\{(v_a, v_b), 1\}$ to $E$ if not already present, do nothing otherwise.
    \item $\omega = \omega^w$ (weighted case): add a new edge $e=\{(v_a, v_b), 1\}$ to $E$ if not already present, otherwise update the existing edge incrementing its weight $w$ by one.
\end{itemize}
We remark that the graph is oriented, which means that $(v_a, v_b)$ and $(v_b, v_a)$ represent two different edges having two different weights.

Since the graph is populated iteratively, given a list of edges $\bm{e} = [ e_0, e_1, e_2, ... ]$, we have that 
\begin{equation}\label{eq:G_i}
G_i = \Lambda(G_{i\text{-}1}, e_i)
\end{equation}
is the intermediate graph resulting after insertion of the $i$-th entry of $\bm{e}$, starting from $G_{\text{-}1}$ (the empty graph).
Finally, given a populated graph $G$ and a node $v_i \in V$, we define an extraction function as
\begin{equation}\label{eq:phi}
\Phi(G, v_i) = \{ f_{i_0}, f_{i_1}, ..., f_{i_{7}} \}
\end{equation}
where $f_{i_k}$ are the graph metrics we consider, listed in Table \ref{tab:features}, computed with respect to $v_i$.

\subsection{Feature extraction}

As mentioned, we start from a data set containing a chronologically ordered list of network connections. Since each connection is a link between two terminals identified by the source and destination IP addresses, each $c_i$ can define an edge which links the vertices $z^s_i$ and $z^d_i$. This means that we can see the data set $D$ as an array of edges and we can use it to iteratively populate a graph according to \eqref{eq:G_i}, which we can rewrite as:
\[
\Lambda\left(G_{i\text{-}1}, \left(z^s_i, z^d_i\right)\right) = G_i = (V_i, E_i),
\]
where $z^s_i, z^d_i \in c_i \in D$. 
Hence, $V_i$ and $E_i$ are populated until the $i$-th connection.

In order to follow the evolution of the graph with good granularity, let us divide the data set into blocks and introduce a parameter ($\sigma \le N$) that defines the size of each block of connections that is progressively used to populate the graph.
This way, the $i$-th connection is associated to the metrics computed on a block-wise populated graph $G_\phi$, where 
\[
\phi = \begin{cases}
    \sigma \cdot \lceil i / \sigma \rceil - 1 & \text{if } \lceil i / \sigma \rceil < \lceil N / \sigma \rceil, \\
    N-1 & \text{otherwise.}
\end{cases}
\]
For example, for $N=129$ and $\sigma=50$, the connection corresponding to $i=63$ will be associated to the metric computed on $G_{99}$, while the connection corresponding to $i=114$ will be associated to the metrics computed on $G_{128}$.
We can consequently update \eqref{eq:phi} into
\begin{equation}
    \Phi^\sigma(z_i) = \Phi(G_\phi, z_i),
\end{equation}
and, for each $c_i \in D$, we can compute the associated graph-based features sets as:
\begin{align}
    \bm{f}^s_i &= \Phi^\sigma(z^s_i), \\
    \bm{f}^d_i &= \Phi^\sigma(z^d_i).
\end{align}

After computing the set of graph-based features for each connection $c_i$, we can generate the new data set $D^*$ as
\begin{equation}
    D^* = [ c^*_0, c^*_1, ..., c^*_{N-1} ],
\end{equation}
where
\begin{equation}
    c^*_i = \{ z^s_i, z^d_i, m_i, \bm{f}^s_i, \bm{f}^d_i \}
\end{equation}
It is important to note that, in the new data set $D^*$, the \textit{classical features} $\bm{a}_i$ are not included, as we no longer consider them in the classification phase.
From now on we will refer to graph metrics sets  $\bm{f}_i$ as \textit{graph features}.


Following the procedure described above, and starting from a data set containing a list of connections, we can generate different new data sets, according to the following criteria:
\begin{itemize}
    \item The parameter $\omega$ can be changed to generate weighted or unweighted graphs. Moreover, we can also consider a \textit{mixed case} (which we denote as $\omega^m$), in which the weight is considered only for the \textit{Degree centrality} related features.
    \item The $\sigma$ parameter has the effect to populate the data set following the evolution of the connections established by each IP address. For example, if $\sigma = N$, we perform graph-based feature extraction only once after collecting all network traffic, such that all connections $c^*$ involving the same network node have the same graph feature set. If $\sigma < N$, we instead perform $\lceil N /\sigma \rceil$ extractions based on $\lceil N /\sigma \rceil$ different graphs, which are progressively populated from connections, following the evolution of each network terminal behaviour. This allows the model to learn on progressive steps of the network behaviour and not only on the final snapshot.
\end{itemize}

\section{Machine Learning-based detection
\label{sec:classification}
}

We consider a binary classification problem using only graph-based features generated according to our approach, with a \textit{Support Vector Machine} algorithm in the non-linear version (\textit{SVM-RBF}). By leveraging the \textit{Radial Basis Function} as the \textit{kernel trick}, we achieve a transformed space in which the data are linearly separable.

For our experiments we consider the \textit{CIC-IDS2017} data set \cite{CIC}. 
It contains network traffic, collected in 5 consecutive days, characterised by 80 traffic-related features extracted with the CICFlowMeter software \cite{CICFlowMeter}. Each connection included in the data set has a label $m$ representing the normal behaviour or the attack category.

Starting from the network traffic data set described above, the first step that must be performed is the generation of the graph-based data set $D^*$, as described in the previous section.
After this, the generated data set must be split into training and testing sets. Considering that the traffic observed during the first day only includes normal behaviour, we use the first two days of traffic for the training set definition, while the test set consists of the traffic observed during the last three days. Specifically, for the training set definition, we adopt the following approach on the first two days of traffic:
\begin{itemize}
    \item We take all the malicious connections of the second day (13835). Such connections belong to \textit{FTP-Patator} and \textit{SSH-Patator}.
    \item We perform an \textit{undersampling} step of the benign class by randomly choosing as many benign samples as the total number of malicious ones chosen in the previous step, and discard the rest.
\end{itemize}

Table \ref{tab:class-distribution} shows the class distribution of the CIC-IDS2017 data set and the sizes of the training and test data sets we created for each class. It turns out that the training set is $0,98\%$ of the entire data set, while the test set is $65,53\%$. It is important to note that the test set was solely used to evaluate the performance of the model, whereas all the training steps, i.e., steps 2) to 5) described next, are performed only on the training set.

\begin{table}[ht]
    \centering
    \caption{CIC-IDS2017 data set and training and testing data sets composition}
    \scalebox{1}{
    \begin{tabular}{@{}llrrrr@{}}
    \toprule
                &                            & \multicolumn{1}{l}{Total} && \multicolumn{1}{l}{Training} & \multicolumn{1}{l}{Test}   \\ 
    \multicolumn{2}{l}{Benign}               &                           &&                              &                            \\
                & \textbf{Total}             & \textbf{2273097}          && \textbf{13835}              & \textbf{1311105}           \\
                & \textbf{\%}                & \textbf{}                 && \textbf{0,61\%}                 & \textbf{57,68\%}              \\[1mm]
    \multicolumn{2}{l}{Malicious}            &                           &&                              &                            \\
                & Bot                        & 1966                      && 0                         & 1966                        \\
                & DDoS                       & 128027                    && 0                        & 128027                      \\
                & DoS GoldenEye              & 10293                     && 0                         & 10293                       \\
                & DoS Hulk                   & 231073                    && 0                        & 231073                     \\
                & DoS Slowhttptest           & 5499                      && 0                         & 5499                       \\
                & DoS slowloris              & 5796                      && 0                         & 5796                       \\
                & FTP-Patator                & 7938                      && 7938                         & 0                       \\
                & Heartbleed                 & 11                        && 0                            & 11                          \\
                & Infiltration               & 36                        && 0                           & 36                         \\
                & PortScan                   & 158930                    && 0                        & 158930                     \\
                & SSH-Patator                & 5897                      && 5897                         & 0                       \\
                & Web Attack - Brute Force   & 1507                      && 0                         & 1507                        \\
                & Web Attack - Sql Injection & 21                        && 0                           & 21                          \\
                & Web Attack - XSS           & 652                       && 0                          & 652                        \\[1mm]
                & \textbf{Total}             & \textbf{557646}           && \textbf{13835}              & \textbf{543811}            \\
                & \textbf{\%}                & \textbf{}                 && \textbf{2,48\%}                & \textbf{97,52\%}              \\[1mm]\midrule
    \multicolumn{2}{l}{Total}                & 2830743                   && 27670                       & 1854916                    \\
    \multicolumn{2}{l}{\% over whole}        &                           && 0,98\%                         & 65,53\%                       \\ \bottomrule
    \end{tabular}
}
\label{tab:class-distribution}
\end{table}

\begin{table*}[h!]
\centering
\caption{Feature selection, hyperparameters tuning, model robustness and model performance results}
\label{tab:train_results}
\scalebox{1}{
\begin{tabular}{@{}llrrrrrrrrrrr@{}} 
\toprule
&                           & \multicolumn{3}{c}{$\omega^u$}                                          &  & \multicolumn{3}{c}{$\omega^w$}                                          &  & \multicolumn{3}{c}{$\omega^m$}                                                           \\
&                           & \mcc{$\sigma^1$}                & \mcc{$\sigma^5$}  & \mcc{$\sigma^N$}  &  & \mcc{$\sigma^1$}                & \mcc{$\sigma^5$}  & \mcc{$\sigma^N$}  &  & \mcc{$\sigma^1$}                & \mcc{$\sigma^5$}                    & \mcc{$\sigma^N$} \\\cmidrule(lr){3-5}\cmidrule(lr){7-9}\cmidrule(lr){11-13}
\multicolumn{5}{l}{\textbf{Feature selection}}                                                                                                                         &  &                                 &                   &                   &  &                                 &                                     &                  \\
& F1          & 0.9999 & 0.9999 & 1 &        & 1 & 1 & 1 &      & 0.9999 & 0.9999 & 1 \\
& \# features & 2       & 2     & 2 &        & 4 & 3 & 2      &      & 2      & 2      & 2 \\
& Most significant feature  & \textit{CC} & \textit{CC} & \textit{CC} &  & \textit{CC} & \textit{CC}       & \textit{DC} &  & \textit{CC} & \textit{CC} & \textit{DC} \\
\multicolumn{5}{l}{\textbf{Hyperparameters Tuning}}                                                                                                                    &  &                                 &                   &                   &  &                                 &                                     &                  \\
& F1 score before tuning    & 0.9999 & 0.9999 & 1 &        & 1 & 1 & 1 &      & 0.9999 & 0.9999 & 1 \\
& F1 score after tuning     & 0.9999 & 0.9999 & 1 &        & 1 & 1 & 1 &      & 0.9999 & 0.9996 & 1 \\
& Percentage increment      & 0.00\% & 0.00\% & 0.00\% &   & 0.00\% & 0.00\% & 0.00\% &    & 0.00\% & -0.3\% & 0.00\% \\
& Chosen \(\gamma\)         & 1 & 1 & 1 &  & 1 & 1 & 0,1 &  & 1 & 1 & 0,1 \\
& Chosen \(C\)              & $10^2$ & $10^2$ & $10$ &  & $1$ & $1$ & 1 &  & $10^2$ & $ 10^2$  & 1 \\
\multicolumn{5}{l}{\textbf{Model robustness}}                                                                                                                          &  &                                 &                   &                   &  &                                 &                                     &                  \\
& F1 avg          & 0.9999 & 0.9999 & 1 &       & 0.9999 & 0.9999 & 1 &        & 0.9999 & 0.9999 & 1 \\
& F1 dev. std     & 0.0001 & 0.0001 & 0 &       & 0.0001 & 0.0001 & 0 &        & 0.0001 & 0.0001 & 0 \\
& Support vectors &   9     &    9  & 27 &      &   93   &   94   & 282  &     &  9  &  9  & 282 \\\midrule
\multicolumn{5}{l}{\textbf{Model performance on test data set}}                                                                                                        &  &                                 &                   &                   &  &                                 &                                     &                  \\
& F1 weighted avg & 
0.9988 & \textbf{0.9988} & 0.9987 &  & 
0.6016 & 0.60 & 0.9988 &  & 
0.9988 & 0.9988 & \textbf{0.9988} \\

& Negatives/Benign          &                                 &                   &                   &  &                                 &                   &                   &  &                                 &                                     &                  \\
& \ \ \ Errors (False Positives) & 
283  & 272  & 439 &  & 
1635 & 1642 & 133 &  & 
283  & 272  & 133 \\
& \ \ \ F1 & 
0.9991 & 0.9991 & 0.9991 &  & 
0.8311 & 0.8306 & 0.9992 &  & 
0.9991 & 0.9991 & 0.9992 \\
& \ \ \ FPR & 
0.0002 & 0.0002 & 0.0003 &  & 
0.0012 & 0.0012 & 0.0001 &  & 
0.0002 & 0.0002 & 0.0001 \\
& Positives/Attacks         &                                 &                   &                   &  &                                 &                   &                   &  &                                 &                                     &                  \\
& \ \ \ Errors (False Negatives) & 
2016   & 2016   & 2016 &  & 
530382 & 532505 & 2016 &  & 
2016   & 2016   & 2016 \\
& \ \ \ F1 & 
0.9979 & 0.9979 & 0.9977 &  & 
0.0480 & 0.0406 & 0.9980 &  & 
0.9979 & 0.9979 & 0.9980 \\
& \ \ \ FNR & 
0.0037 & 0.0037 & 0.0037 &  & 
0.9753 & 0.9792 & 0.0037 &  & 
0.0037 & 0.0037 & 0.0037 \\


\bottomrule
\end{tabular}
}
\end{table*}

After having generated the graph-based data set as described above, the following subsequent steps need to be performed in order to use it for the purposes of our work. From now on with \textit{model} we mean the SVM-based ML classifier, and with \textit{model configuration} we denote the training parameter set and the feature selection process.
\begin{enumerate}
    
    \item Features Scaling: the technique used involves standardisation based on the means and standard deviations computed from the training set, and then applied to both the training and the test sets. This procedure has only been performed on features with an unrestricted domain, namely \textit{Degree centrality} and its variants (Table \ref{tab:features}).
    \item Feature Selection: we consider \textit{Forward Feature Selection} (FFS) with a \textit{SVM-RBF} estimator to find a space in which data is successfully separable. A maximum of eight features has been chosen as an upper bound. Selections are not pursued if F1 score saturation is reached. The results were validated through a 5-fold cross-validation.
    \item Hyperparameters tuning: the \textit{SVM-RBF} requires tuning of the hyperparameters \textit{\(\gamma\)} and \textit{\(C\)}, which must be chosen accurately. We optimized them through a 5-fold cross-validated \textit{grid search} with
        \begin{itemize}
            \item \(C = \{0.1, \, 1,\, 5,\, 10,\, 10^2,\, 10^3,\, 10^4,\, 10^5\}\)
            \item \(\gamma = \{0.01,\, 0.1,\, 0.5,\, 1\}\)
        \end{itemize}
    \item Model robustness: once the best model configuration has been found, 10-fold cross-validation has been used for calculating, for each fold, the performance metrics of the fold-trained model. In particular, we look at the F1 score standard deviation, which provides an insight into the consistency of the model's performance and its ability to generalise to unseen data. A lower standard deviation indicates a more stable and reliable model, which corresponds to our case as shown in Table \ref{tab:train_results}.
    \item Model training: we perform training on the whole training set. The resulting model is then used for performing tests on the test data set. The classification performance, so obtained, is reported in Table \ref{tab:train_results}.
\end{enumerate}

The above process has been repeated for each combination of the following two parameter sets:
\begin{itemize}
    \item $\omega$ parameter set: $\{ \omega^w, \omega^w, \omega^m\}$.
    \item $\sigma$ parameter set: $\{ \sigma^1, \sigma^5, \sigma^N \}$.
\end{itemize}

\section{Results and discussion
\label{sec:results}
}

The approach and techniques described in the previous sections have been validated through numerical simulations.
In order to ensure maximum reproducibility of our experiments, the source code used to perform the simulations has been made publicly available\footnote{\url{https://github.com/secomms/GiBIDS}}.
Based on the results of our experiments, which are reported in Table \ref{tab:train_results}, the following considerations are in order.

\subsubsection{Feature Selection}

We observe that, except for the \textit{weighted} case with $\{\sigma^1, \sigma^5\}$, always two features are selected. Such features allow to obtain an almost perfect separation of training data instances ($F1 \geq 0.9999$).
In all cases, the \textit{Clustering Coefficient} (\textit{CC}) is the most significant feature,  i.e., the most capable measure to separate training data, except for the \textit{weighted} and \textit{mixed cases} with $\sigma=N$, for which such a role is played by the \textit{Degree Centrality} (\textit{DC}).

\subsubsection{Hyperparameters tuning}

Having achieved an almost perfect separation of the training data instances, tuning the model hyperparameters doesn't produce any significant improvement in the F1 score. Nevertheless, such a step is important for minimising the complexity of the hyperplane. We observe that for the hyperparameter $\gamma$ the largest possible value is chosen, except for the \textit{weighted} and \textit{mixed cases} with $\sigma=N$. For the hyperparameter \(C\), small values are always chosen ($C \leq 10^2$). Such results show that the training instances are already well separated and do not require a complex hyperplane definition. In particular in the \textit{weighted} and \textit{mixed cases} cases with $\sigma=N$, the minimum value is chosen for both hyperparameters.

\subsubsection{Model robustness}

We observe that the average F1 score is never smaller than $0.9999$ and that the standard deviation is never greater than $0.0001$, regardless of the values of $\sigma$ and $\omega$, demonstrating good robustness of the model performance. For what concerns model complexity, we note that the number of \textit{support vectors} chosen by the \textit{SVM-RBF} is never greater than $1\%$ of the training set, ensuring a low model complexity.

\subsection{Model performance on test data set}

Performance achieved on test data is reported at the bottom of Table \ref{tab:train_results}, for all the considered combinations of $\sigma$ and $\omega$. 
We observe that $\omega^u$ and $\omega^m$ achieve the same results for $\sigma<N$, while $\omega^w$ and $\omega^m$ achieve the same results for $\sigma=N$. This can be explained by the fact that for $\sigma^1$ and $\sigma^5$ the selected features (CC and BC) do not depend on the weights, while for $\sigma^N$ they (DC) do. 
For the case with $\sigma=N$, we observe that classifiers working on the weighted graph ($\omega^w$) are unable to distinguish normal from malicious behaviour. This suggests that 
it must hold that $\omega \neq \omega^w$, because although the classifier is able to separate training set instances, it fails with instances from the test set, resulting in overfitting. For this reason, we discard the \textit{weighted case}. 
Since the misclassified attacks are always the same for both $\omega^u$ and $\omega^m$ (all \textit{Bot}, \textit{Heartbleed} and \textit{Infiltration} attacks and three \textit{DDoS} attacks), also the \textit{False Negative Rate} (FNR) is always the same. Therefore, we must base the decision of the best model for $\sigma<N$ and $\sigma=N$ on the \textit{False Positive Rate} (FPR). This leads to the selection of $\{\sigma^5, \omega^u\}$ ($FPR=0.02\%$) and $\{\sigma^N, \omega^m\}$ ($FPR=0.01\%$) as the best configurations. These models not only achieve excellent detection performance for $99.63\%$ of the attacks tested, failing only for the attacks mentioned above, but also manage to maintain a very low FPR.

\subsection{Comparison with previous approaches}

By comparing our approach with the state of the art, as done in Table \ref{tab:results}, it can be seen that with a smaller number of features required, better results are achieved in every evaluation metric considered. It should also be noted that the size of the training set is significantly smaller than that of other approaches, highlighting both the model ability to achieve better detection performance despite a smaller number of training samples and the reliability of the results obtained by testing the model on a larger test data set.

\begin{table}[t]
\caption{Comparison with the state of the art }
\label{tab:results}
\scalebox{0.95}{
\begin{tabular}{lrrrr}
    \toprule
    \textbf{IDS Model}                     & \textbf{\% train-test}  & \textbf{\textbf{\# ftrs}}  & \textbf{F1}  & \textbf{FPR} \\\midrule
Sharafaldin et al. \cite{CIC}          & N.A.                    & 49                         & 0.98         & N.A.         \\
Abdulhammed et al. \cite{Abdulhammed}  & 70-30                   & 10                         & 0.997        & 0.1\%        \\
Rodriguez et al. \cite{rodriguez}      & 50-50                   & 6                          & 0.990        & N.A.         \\
Thockchom et al. \cite{Thockchom}      & 80-20                   & 13                         & 0.9963       & 0.43\%       \\
Our approach $\{\sigma^5,\omega^u\}$      & 0.98-65.53                   & 2                          & 0.9988       & 0.02\%       \\
Our approach $\{\sigma^N,\omega^m\}$         & 0.98-65.53                   & 2                          & 0.9988       & 0.01\%       \\

    \bottomrule
    \end{tabular}
    }
\end{table}

\section{Conclusion
\label{sec:conclusion}
}

We proposed the use of metrics derived from graph theory to detect cyber threats from network traffic.
By using a graph to model interactions between network nodes, instead of analysing the content of the information exchanged between them, we are able to overcome some known limitations of previous approaches, like difficulty of analyzing encrypted packets and of detecting malicious behaviours when attackers use elusive techniques.
We validated our approach through experiments considering the CIC-IDS2017 data set, showing that it is indeed capable of achieving better results than previous approaches, which analyse each connection as a stand-alone entity whereby the evolution of the involved terminals' behaviour with the rest of the network is not considered.
These results are promising and suggest further research, for example by executing a larger campaign of experiments on diverse scenarios.

\bibliographystyle{IEEEtran}
\bibliography{sample}

\end{document}